\documentclass{elsart4-1}
  \usepackage{graphicx,color}
  \usepackage{epsfig}
  \usepackage{amssymb}
  \usepackage[english,francais]{babel}
  \definecolor{Modif}{rgb}{0.0,0.0,0.0}

\renewcommand{\theequation}{\arabic{equation}}
\def\og{\leavevmode\raise.3ex\hbox{$\scriptscriptstyle\langle\!\langle$~}}
\def\fg{\leavevmode\raise.3ex\hbox{~$\!\scriptscriptstyle\,\rangle\!\rangle$}}

\begin{document}
% Choisir la rubrique principale 'Physique' ou 'Astrophysique', et
%      apres une rubrique
\centerline{Astrophysique/Techniques astronomiques}

\begin{frontmatter}

 \selectlanguage{francais}
 \title{Un coronographe interf{\'e}rentiel achromatique coaxial}

 \author[Nice]{Jean Gay},
 %\ead{Jean.Gay@obs-nice.fr}
 \author[Nice]{Fran{\c c}ois Fressin},
 %\ead{Francois.Fressin@obs-nice.fr}
 \author[Nice]{Jean-Pierre Rivet\thanksref{CorrAuth}},
 \thanks[CorrAuth]{Corresponding author}
 \ead{Jean-Pierre.Rivet@obs-nice.fr}
 \author[Roq]{Yves Rabbia},
 %\ead{Yves.Rabbia@obs-azur.fr}
 \author[Roq]{Christophe Buisset}
 %\ead{Christophe.Buisset@obs-azur.fr}

 \address[Nice]{Observatoire de la C{\^o}te d'Azur, B.P. 4229,
 06304 Nice cedex 4, France}
 \address[Roq]{Observatoire de la C{\^o}te d'Azur, Avenue Nicolas Copernic
 06130 Grasse, France}

 % Vous pouvez ajouter a la prochain ligne les dates
 %   de reception et d'acceptation, et apres le nom du presentateur
 \medskip
 \begin{center} \small
 %Soumis le 12 juillet 2005,\\
  Re\c{c}u le 20 juillet 2005; accept{\'e} le $1^{er}$ d{\'e}cembre 2005\\
  Disponible sur Internet le 5 janvier 2006\\
  pr{\'e}sent{\'e} par Pierre Encrenaz
 \end{center}

 \begin{abstract}
 \selectlanguage{francais}
 Dans le but de parvenir {\`a} l'imagerie {\`a} haute dynamique
 d'objets comme les exoplan{\`e}tes, nous pr{\'e}sentons ici un nouveau
concept de coronographe stellaire interf{\'e}rentiel, le ``CIAXE''. Il
est d{\'e}riv{\'e} du ``CIA'', le Coronographe Interf{\'e}rentiel Achromatique
\cite{GR96a,BRG00a}. Le CIAXE se distingue de son pr{\'e}d{\'e}cesseur par
une combinaison optique originale, simplifi{\'e}e, tr{\`e}s compacte et
totalement coaxiale. En effet, {\`a} la diff{\'e}rence du CIA classique
qui est d{\'e}riv{\'e} de l'interf{\'e}rom{\`e}tre de Michelson, le CIAXE d{\'e}livre
son faisceau de sortie sur le m{\^e}me axe que le faisceau d'entr{\'e}e,
ce qui facilitera grandement son insertion au sein de
l'instrumentation focale d'un t{\'e}lescope. Un tel dispositif
pourrait constituer une avanc{\'e}e en mati{\`e}re d'instrumentation
focale pour la recherche d'exoplan{\`e}tes.
 {\it Pour citer cet article~: J. Gay et al, C. R. Physique 6 (2005).}

 \vskip 0.5\baselineskip
 \selectlanguage{english}
 \noindent{\bf Abstract}
 \vskip 0.5\baselineskip
 \noindent {\bf On-axis achromatic interfero-coronagraph. }
 We present a new type of stellar interfero-coronagraph, the
``CIAXE'', which is a variant of the ``AIC'', the Achromatic
Interfero-Coronagraph \cite{GR96a,BRG00a}. The CIAXE is
characterized by a very simple, compact and fully coaxial optical
combination. Indeed, contrarily to the classical AIC which has a
Michelson interferometer structure, the CIAXE delivers its output
beam on the same axis as the input beam. This will ease its
insertion in the focal instrumentation of existing telescopes or
next generation ones. Such a device could be a step forward in the
field of instrumental search for exoplanets.
 {\it To cite this article: J. Gay et al, C. R. Physique 6 (2005).}

 \vskip 0.5\baselineskip
 \noindent
 {\small{\it Mots-cl{\'e}s~:} Coronographie stellaire~; Exo-plan{\`e}tes; Imagerie {\`a} haute dynamique}}
\end{abstract}

\end{frontmatter}

            \selectlanguage{english}
            \section*{Abridged English version}

The detection of faint objects (like exoplanets or protoplanetary
disks) near bright astrophysical sources requires astronomical
instruments with both a high angular resolution and a high
dynamical photometric range. An efficient solution to achieve high
dynamics imaging is to use a stellar coronagraph. Such devices are
intended to reject most of the light from the bright source
on-axis, in order to make a faint off-axis source detectable. This
concept has been first introduced in 1931 by B.~Lyot, to observe
the solar corona. Since, several stellar coronagraphs have been
proposed, such as the phase mask coronagraph \cite{RR97a}, or the
sectorized mask coronagraph \cite{RRBCL00a}.

 In 1996, J.~Gay and Y.~Rabbia \cite{GR96a,BRG00a} have introduced a
new concept of stellar coronagraph, the AIC (Achromatic
Interfero-Coronagraph), involving interferential rejection by
amplitude division. It was based on a Michelson interferometer
structure, but with one of its arms replaced by a ``cat's eye''.
The latter has the property of reversing the pupil and provides an
achromatic $\pi$ phase shift by focus crossing \cite{G90a,BW99a}.
The beam splitter of the Michelson structure separates the
incoming parallel light beam into two beams. One of them travels
along the ``active'' arm of the interferometer (the one with the
cat's eye) and the other travels along the ``passive'' one (with
flat mirrors only). The first beam undergoes a pupil reversal and
$\pi$ phase shift, whereas the second does not. So, when they
recombine, the light coming from an object lying on the axis (thus
invariant by pupil reversal), is rejected by destructive
interference. With a perfect input wavefront and perfect optical
components, the rejection rate would be infinite. Besides, the
light from an off-axis object is \emph{not} invariant by pupil
reversal and thus does \emph{not} vanish by interference.

The standard AIC itself was an assembly of ten optical components.
In addition, due to its Michelson-like structure, the output beam
of the AIC is delivered at right angle from the input beam. This
requires extra optical components for a proper insertion in an
existing (coaxial) optical train. To circumvent this problem, a
fully on-axis version has been designed. It lies upon the same
principle as the AIC, but involves only two coaxial thick lenses
machined in the same optical material. A thin, regular and
carefully calibrated air gap between the two lenses acts as a beam
splitter. Both lenses must have equal thickness $e$, in order to
guarantee a zero optical paths difference within the
interferometer.

 The optical combination is described in figure~\ref{fig:CIAXE_AB}. The
beam-splitter surface $M_2$ between the two thick lenses is drawn
as a single thick dashed line, albeit it involves actually two
surfaces with equal curvature radii $R_2$, separated by a thin
spacer ring. The surfaces $M_1$ and $M_3$ are coated so as to be
fully reflecting, except for small circular areas surrounding
$S_1$ and $S_3$. These areas act as input and output field stops.

The CIAXE must be fitted in the focal plane of a telescope, so
that the incident beam coming from an on-axis source converge onto
$S_1$, at the center of the input aperture. Then, it reaches the
beam splitter $M_2$. It is divided into a reflected beam ``$A$''
(thin dashed lines on Figure~\ref{fig:CIAXE_AB}) and a transmitted
beam``$B$'' (thin dotted lines on Figure~\ref{fig:CIAXE_AB}). The
reflected beam $A$ reaches $M_1$, where it is reflected again so
as to converge onto $S_3$ (through $M_2$). The transmitted beam
$B$ reaches $M3$ where it is reflected so as to converge onto a
real ``focus'' $F$, then reaches back $M_2$, which reflects it, so
as to make it converge onto $S_3$ at the center of the output
aperture. Beam $B$ crosses the real focal point ($F$) and thus
undergoes a $\pi$ phase shift and a pupil reversal, whereas
Beam~$A$ does not. Provided the optical lengths be equal, both
beams converging onto $S_3$ vanish by destructive interferences.
So, the light of an on-axis object is completely rejected at the
output, at least for a perfect device and a perfect input
wavefront, in the limit of low incidence angles.

Secondary beams that are reflected/transmitted several times by
$M_2$ do exist, but it can be shown that they merely send the
energy of an on-axis source back to the input.

 The optical
combination which fulfills all the required constraints is unique
(up to a symmetry and a scaling factor) and displays a range of
noticeable optical properties.

Within the limit of vanishing numerical aperture of the input beam
(or with correct spherical aberration rejection), the performances
are expected to be identical to those of the original AIC
\cite{BRG00a}. This device combines the advantages of the original
AIC with the advantages of a fully coaxial, compact and
phase-locked device. It is thus a potential candidate for
exoplanets research from ground-based or space-based telescopes.

            \selectlanguage{francais}
            \section{Introduction}
            \label{sec:Intro}

 L'{\'e}tude par imagerie d'objets astronomiques
t{\'e}nus au voisinage d'objets tr{\`e}s brillants (par exemple une
exo-plan{\`e}te, une naine brune, un disque d'accr{\'e}tion ou un disque
proto-plan{\'e}taire) n{\'e}cessite des moyens d'observation qui combinent
la haute r{\'e}solution angulaire, et la haute dynamique
photom{\'e}trique. Une m{\'e}thode prometteuse pour obtenir cette haute
dynamique est d'avoir recours {\`a} un coronographe stellaire. Le
terme ``coronographe'' est employ{\'e} ici par analogie avec le
coronographe solaire con{\c c}u par Bernard Lyot en 1931 pour
observer la couronne solaire. Depuis, plusieurs types de
coronographes stellaires ont {\'e}t{\'e} imagin{\'e}s. Ce genre d'instrument
focal s'intercale entre le t{\'e}lescope collecteur de lumi{\`e}re et la
camera d'imagerie. Il att{\'e}nue fortement la lumi{\`e}re de la source
tr{\`e}s lumineuse ({\'e}toile) situ{\'e}e au centre du champ, pour rendre la
lumi{\`e}re t{\'e}nue du compagnon plus facilement d{\'e}tectable. Citons par
exemple le coronographe {\`a} masque de phase de C.~et F.~Roddier
\cite{RR97a}, ou le coronographe {\`a} quatre quadrants de D.~Rouan
{\sl et al.} \cite{RRBCL00a}.

En 1996, J.~Gay et Y.~Rabbia \cite{GR96a,BRG00a} ont introduit un
nouveau concept de coronographe stellaire: le CIA (Coronographe
Interf{\'e}rentiel Achromatique), qui assure l'att{\'e}nuation de l'objet
central par des interf{\'e}rences destructives apr{\`e}s division de flux.
Il se distingue des autres mod{\`e}les de coronographes par le fait
qu'il est achromatique et qu'il n'introduit pas de discontinuit{\'e}
dans les surfaces d'onde (il proc{\`e}de par division de flux, et non
par division de front d'onde). Il poss{\`e}de une configuration
d{\'e}riv{\'e}e de l'interf{\'e}rom{\`e}tre de Michelson, {\`a} ceci pr{\`e}s que l'un des
bras comporte un dispositif en ``\oe il de chat''. Ce dernier a la
propri{\'e}t{\'e}
\textcolor{Modif}{d'effectuer une sym{\'e}trie centrale sur la pupille},        %@@@@1
et un d{\'e}phasage achromatique de $\pi$ de l'amplitude (cette propri{\'e}t{\'e}
de d{\'e}phasage
de $\pi$ pour un faisceau lumineux qui traverse un point focal
r{\'e}el est bien connue depuis le \emph{XIX}$^{e}$ si{\`e}cle; voir par
exemple les travaux de L.G.~Gouy (1890) \cite{G90a} ou le livre de
Born et Wolf \cite{BW99a}). La s{\'e}paratrice de l'interf{\'e}rom{\`e}tre
divise le faisceau parall{\`e}le d'entr{\'e}e en deux faisceaux
d'intensit{\'e}s {\'e}gales (id{\'e}alement). L'un de ces faisceaux emprunte
le bras ``actif'' de l'interf{\'e}rom{\`e}tre (celui comportant l'\oe il
de chat), et le second emprunte le bras ``passif'' (ne comportant
que des miroirs plans). Le premier faisceau subit donc
\textcolor{Modif}{une sym{\'e}trie centrale de sa pupille}                      %@@@@1
et un d{\'e}phasage achromatique de $\pi$, ce 
qui ne se produit pas pour le second. De ce fait, lorsque les deux
faisceaux se recombinent, la lumi{\`e}re d'un objet non r{\'e}solu situ{\'e}
sur l'axe optique (donc invariant par
\textcolor{Modif}{l'op{\'e}ration de sym{\'e}trie centrale})                        %@@@@1
est fortement att{\'e}nu{\'e}e par interf{\'e}rence destructive. Avec un front
d'onde incident parfaitement plan et une optique parfaite, la
r{\'e}jection de la lumi{\`e}re incidente d'un objet sur l'axe serait
totale. En revanche, pour un objet situ{\'e} hors-axe, donc
non-invariant par
\textcolor{Modif}{l'op{\'e}ration de sym{\'e}trie centrale},                       %@@@@1
l'annulation interf{\'e}rom{\'e}trique ne se produit pas.

Le CIA ``classique'' est lui-m{\^e}me un assemblage de 10
composants optiques. De plus, {\`a} cause de sa structure h{\'e}rit{\'e}e de
l'interf{\'e}rom{\`e}tre de Michelson, il d{\'e}livre son faisceau de sortie {\`a}
angle droit par rapport {\`a} son faisceau d'entr{\'e}e, ce qui impose
l'introduction d'une optique d'adaptation externe pour l'insertion
d'un CIA dans le train optique (habituellement coaxial) d'un
ensemble t{\'e}lescope-cam{\'e}ra. C'est pour contourner cette difficult{\'e}
qu'a {\'e}t{\'e} con{\c c}u le ``CIAXE'', une version simplifi{\'e}e du CIA
(deux composants optiques seulement), compl{\`e}tement coaxiale, dont
nous livrons ci-apr{\`e}s la description.

          \section{Le principe optique}
          \label{sec:Principe}

 Le CIAXE repose sur un jeu de deux lentilles
{\'e}paisses, taill{\'e}es toutes les deux dans le m{\^e}me mat{\'e}riau optique,
et dont les {\'e}paisseurs au centre sont identiques. Ces deux
lentilles sont assembl{\'e}es de mani{\`e}re coaxiale, avec une fine
entretoise entre les deux, de mani{\`e}re {\`a} m{\'e}nager une lame d'air
d'{\'e}paisseur uniforme et calibr{\'e}e. Cette couche d'air servira de
lame s{\'e}paratrice sym{\'e}trique, pour diviser le faisceau incident en
deux parties {\'e}gales en {\'e}nergie. L'utilisation d'un mat{\'e}riau {\`a} fort
indice de r{\'e}fraction, comme le $ZnSe$, permet de r{\'e}aliser ainsi
une s{\'e}paratrice {\'e}quilibr{\'e}e sans aucun d{\'e}p{\^o}t m{\'e}tallique. 
\textcolor{Modif}{Par exemple, avec une {\'e}paisseur d'air de $0,55\,\mu m$,                %@@@@2
la s{\'e}paratrice sera {\'e}quilibr{\'e}e (efficacit\'e $4RT$ sup{\'e}rieure {\`a} $0,9$) entre
$1,46\,\mu m$ et $4,26\,\mu m$ pour des incidences inf{\'e}rieures {\`a} $3^o$,
toutes polarisations confondues.}          
L'{\'e}galit{\'e} des {\'e}paisseurs au centre des deux lentilles garantit que les
longueurs des deux chemins optiques de l'interf{\'e}rom{\`e}tre soient
identiques par construction,
\textcolor{Modif}{pour toute longueur d'onde, malgr{\'e} la dispersion du mat{\'e}riau}.         %@@@@3
Aucun r{\'e}glage fastidieux et instable
de la diff{\'e}rence de marche ne sera donc n{\'e}cessaire.

\begin{figure}[!h]
\begin{center}
\epsfig{file=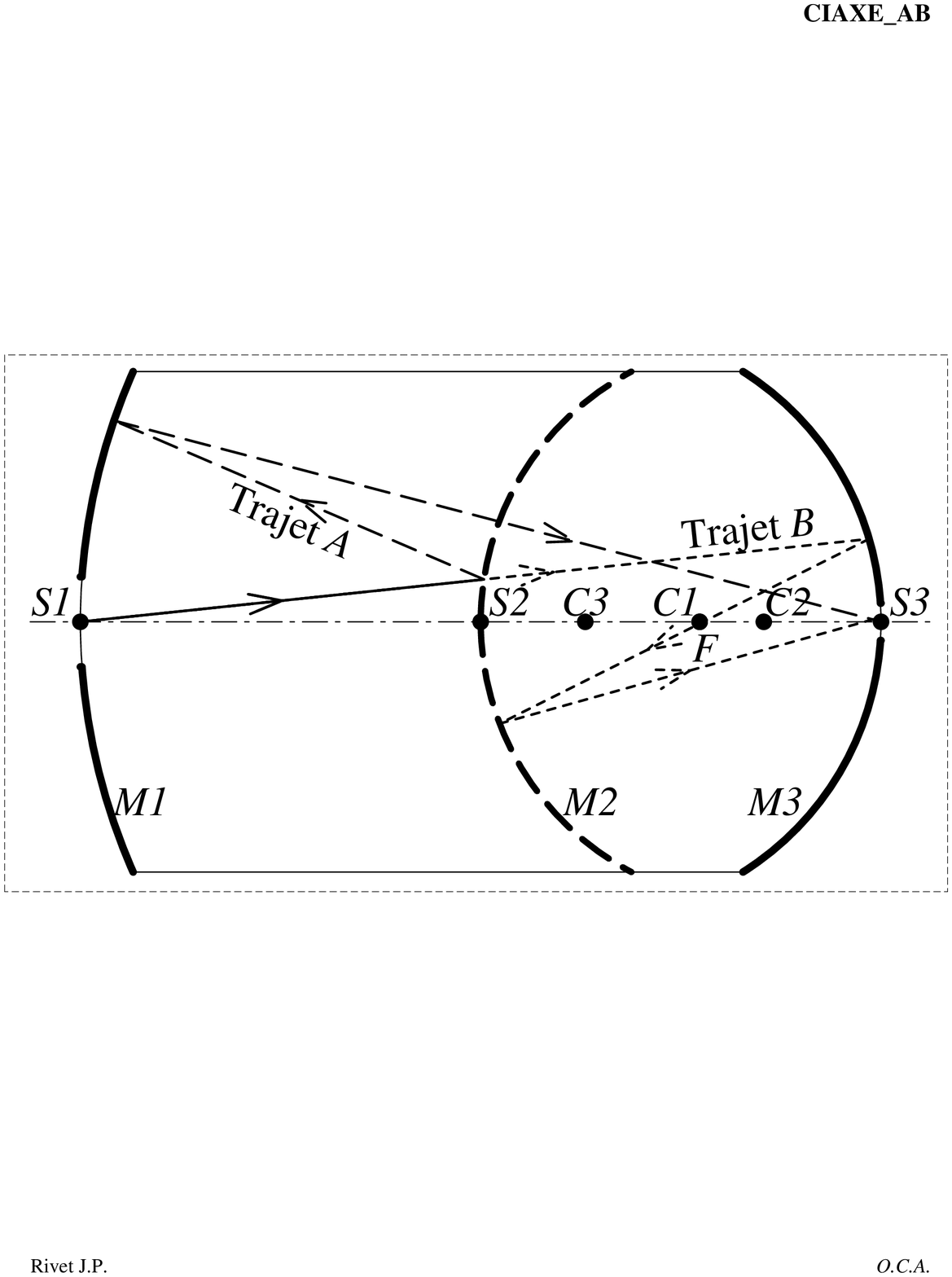,width=6cm,angle=0,clip=}
\end{center}
\caption[Le CIAXE]{\small\it Le principe optique du CIAXE. Les
surfaces trait{\'e}es r{\'e}fl{\'e}chissantes sont en trait fort continu. la
surface s{\'e}paratrice semi-r{\'e}fl{\'e}chissante est en trait tiret{\'e} fort.
Le rayon $A$, en trait tiret{\'e} fin, est r{\'e}fl{\'e}chi par $M_2$, et
atteint l'ouverture de sortie $S_3$, sans passer par un point
focal r{\'e}el. Le rayon $B$, en trait pointill{\'e} fin, est transmis par
$M_2$, et passe par le point focal $F$ avant d'atteindre
l'ouverture de sortie.}
 \label{fig:CIAXE_AB}
\end{figure}

La combinaison optique est d{\'e}crite en Figure~\ref{fig:CIAXE_AB}.
Par souci de clart{\'e}, la s{\'e}paratrice $M_2$ entre les deux lentilles
y est repr{\'e}sent{\'e}e comme un simple trait tiret{\'e} {\'e}pais, bien qu'elle
soit en r{\'e}alit{\'e} constitu{\'e}e par deux surfaces de m{\^e}me rayon de
courbure $R_2$, s{\'e}par{\'e}es par une fine lame d'air d'{\'e}paisseur
calibr{\'e}e. La surface optique $M_1$ sera trait{\'e}e de mani{\`e}re {\`a} {\^e}tre
totalement r{\'e}fl{\'e}chissante (trait continu fort sur la
figure~\ref{fig:CIAXE_AB}), sauf sur une petite surface circulaire
de diam{\`e}tre $d_1$, centr{\'e}e sur le sommet $S_1$, qui servira de
diaphragme de champ (trait continu fin sur la
figure~\ref{fig:CIAXE_AB}). Il en sera de m{\^e}me pour la surface
optique $M_3$, qui portera l'ouverture de sortie de diam{\`e}tre
$d_3$, centr{\'e}e sur $S_3$.
\textcolor{Modif}{Ces deux zones pourront}                                             %@@@@4
{\'e}ventuellement {\^e}tre trait{\'e}es anti-reflet.

Le CIAXE doit {\^e}tre install{\'e} dans le plan focal d'un t{\'e}lescope dont
la pupille, avec son obstruction centrale, doit poss{\'e}der la
sym{\'e}trie centrale autour de l'axe optique du t{\'e}lescope\footnote{En
cas de pupille d'entr{\'e}e non sym{\'e}trique, un masque sym{\'e}trisant doit
{\^e}tre introduit dans un plan pupille interm{\'e}diaire.}. Le CIAXE
sera install{\'e} de fa{\c con} {\`a} ce que le faisceau d{\'e}livr{\'e} par le
t{\'e}lescope pour une {\'e}toile sur l'axe converge en $S_1$, le centre
de l'ouverture d'entr{\'e}e. Ce faisceau atteint ensuite la surface
s{\'e}paratrice, qui le s{\'e}pare en un faisceau r{\'e}fl{\'e}chi (trajet $A$; en
trait tiret{\'e} fin sur la figure~\ref{fig:CIAXE_AB}), et un faisceau
transmis (trajet $B$; en trait pointill{\'e} fin sur la
figure~\ref{fig:CIAXE_AB}).

 Le faisceau r{\'e}fl{\'e}chi $A$ atteint en
suite la surface $M_1$, qui le r{\'e}fl{\'e}chit vers $S_3$, le centre de
l'ouverture de sortie, au sommet de la surface $M_3$. Au passage,
ce faisceau traverse la s{\'e}paratrice $M_2$, ce qui donne lieu {\`a} une
onde r{\'e}fl{\'e}chie secondaire que l'on {\'e}tudiera en
Section~\ref{sec:Refl}.

Le faisceau transmis $B$ atteint la surface  $M_3$, qui le
r{\'e}fl{\'e}chit et le fait converger sur un point focal r{\'e}el $F$. Au
del{\`a} de $F$, le faisceau diverge vers la surface s{\'e}paratrice
$M_2$, qui le r{\'e}fl{\'e}chit vers $S_3$. Ce passage sur $M_2$ cr{\'e}e
{\'e}galement une onde transmise secondaire que l'on {\'e}tudiera aussi en
Section~\ref{sec:Refl}.

Les deux faisceaux ont subi une r{\'e}flexion et une transmission sur
la surface s{\'e}paratrice $M_2$. Le faisceau $B$ a subi une
\textcolor{Modif}{sym{\'e}trie centrale de sa pupille}                                      %@@@@1
(grandissement angulaire oppos{\'e} {\`a} celui du
faisceau $A$) et un passage par le point focal r{\'e}el $F$, qui
impose {\`a} son amplitude un d{\'e}phasage achromatique de $\pi$
\cite{G90a,BW99a}. Le faisceau $A$ {\'e}chappe {\`a} ces deux effets. De
ce fait, si les longueurs des deux trajets optiques sont {\'e}gales et
si la pupille a bien la sym{\'e}trie axiale, la lumi{\`e}re d'une source
situ{\'e}e sur l'axe du CIAXE est compl{\`e}tement {\'e}teinte en sortie, au
moins pour un dispositif id{\'e}alement r{\'e}alis{\'e}, pour une onde
incidente parfaite (absence d'aberrations impaires d'origine
atmosph{\'e}rique ou instrumentale), et dans la limite de l'optique
g{\'e}om{\'e}trique paraxiale (faibles angles d'ouverture).

En revanche, pour une source situ{\'e}e hors de l'axe, la surface
d'onde produite est inclin{\'e}e, et donc non invariante par
\textcolor{Modif}{la sym{\'e}trie centrale}.                                               %@@@@1
L'annulation interf{\'e}rom{\'e}trique ne se
produit plus, et les deux images issues des trajets $A$ et $B$ ne
se superposent plus. Elles donnent deux images sym{\'e}triques de part
et d'autre de l'axe, comme dans le cas du CIA classique. L'effet
coronographique r{\'e}sulte de l'annulation (ou att{\'e}nuation) s{\'e}lective
de la lumi{\`e}re de la source situ{\'e}e sur l'axe, mais pas des sources
hors axe.

 Sous les
conditions mentionn{\'e}es ci-dessus, le fonctionnement
interf{\'e}rom{\'e}trique du CIAXE est identique {\`a} celui d'un CIA
classique, et les {\'e}valuations th{\'e}oriques de performance {\'e}tablies
pour ce dernier \cite{GR96a,BRG00a}, restent valides.

          \section{Les dimensions du dispositif}
          \label{sec:Dimen}

Les dimensions {\`a} fixer sont l'{\'e}paisseur commune $e$ des deux
lentilles, les trois rayons de courbure $R_1$, $R_2$ et $R_3$, le
diam{\`e}tre $D$ de l'ensemble, ainsi que les diam{\`e}tres $d_1$ et $d_3$
des ouvertures d'entr{\'e}e et de sortie. L'{\'e}paisseur de la lame d'air
servant de s{\'e}paratrice semi-r{\'e}fl{\'e}chissante fait l'objet d'un
calcul usuel d'optique des couches minces, qui ne sera pas {\'e}voqu{\'e}
ici.

 Si l'on suppose connue
l'{\'e}paisseur commune $e$ des deux lentilles, alors les rayons de
courbure $R_1$, $R_2$ et $R_3$ des trois surfaces
sph{\'e}riques\footnote{Pour des performances optimales, la surface
$M_3$ doit {\^e}tre en fait l{\'e}g{\`e}rement asph{\'e}rique, avec un coefficient
de conicit{\'e} de $0,6$.} sont compl{\`e}tement d{\'e}termin{\'e}s par les trois
contraintes ci-dessous\,:
 \begin{itemize} \itemsep=0pt
   \item $S_3$ doit {\^e}tre le point conjugu{\'e} de $S_1$ selon le trajet optique $A$.
   \item $S_3$ doit {\^e}tre le point conjugu{\'e} de $S_1$ selon le trajet optique $B$.
   \item Le grandissement angulaire global $\Gamma_A$ pour le faisceau $A$ doit {\^e}tre
   l'oppos{\'e} du grandissement angulaire global $\Gamma_B$ pour le faisceau $B$.
 \end{itemize}
Ce probl{\`e}me poss{\`e}de une solution unique ({\`a} une sym{\'e}trie pr{\`e}s)\,:
 \begin{equation}
   R_1=e\frac{4}{4-\sqrt{2}},\quad
   R_2=e\frac{1}{\sqrt{2}}, \quad
   R_3=e\frac{-4}{4+\sqrt{2}},
 \label{eq:RRR1}
 \end{equation}
o{\`u} $e$ est l'{\'e}paisseur commune des deux lentilles.

Il est important de noter que l'image (virtuelle) de $S_1$ par
r{\'e}flexion sur $M_2$ co{\"\i}ncide avec $C_3$, le centre de courbure de
la surface $M_3$. De m{\^e}me, le point focal r{\'e}el $F$, qui n'est
autre que l'image r{\'e}elle de $S_1$ par r{\'e}flexion sur $M_3$,
co{\"\i}ncide avec $C_1$, le centre de courbure de la surface $M_1$.

Le diam{\`e}tre externe $D$ du CIAXE est fix{\'e} en fonction de
l'{\'e}paisseur $e$ par le rapport d'ouverture du t{\'e}lescope
collecteur. En effet, pour que les divers faisceaux lumineux ne
subissent aucun vignetage {\`a} l'int{\'e}rieur de l'optique du CIAXE, le
rapport $D/e$ doit v{\'e}rifier\,:
 \begin{equation}
  \frac{D}{e}>\frac{\Theta}{n(\lambda)}\,\, \frac{2\sqrt{2}}{\sqrt{2}-1},
  \label{eq:Diam}
 \end{equation}
 o{\`u} $\Theta$ est le rapport d'ouverture du t{\'e}lescope (suppos{\'e}
 petit), et $n(\lambda)$ est l'indice du mat{\'e}riau optique
 constituant le CIAXE. En pratique, cette in{\'e}galit{\'e} est peu contraignante,
 et ce sont des consid{\'e}rations techniques li{\'e}es {\`a} la technique
 d'usinage qui fixent le choix de $D$.

 Les diam{\`e}tres des ouvertures d'entr{\'e}e et de sortie sont fix{\'e}s en fonction de
 l'{\'e}paisseur $e$ par la contrainte que l'une doit {\^e}tre l'image
 de l'autre par le syst{\`e}me complet, et que les deux soient dans
 l'ombre port{\'e}e de l'obstruction centrale du t{\'e}lescope. Ceci
 impose que\,:
 \begin{equation}
  \frac{d_3}{e}<\frac{2\theta}{n(\lambda)}\quad\hbox{et}\quad
  d_1=\frac{d_3}{(\sqrt{2}-1)},
  \label{eq:Diam}
 \end{equation}
o{\`u} $\theta$ est le rapport ``d'ouverture'' (diam{\`e}tre sur focale)
associ{\'e} {\`a} l'obstruction centrale du t{\'e}lescope. Afin de limiter les
effets de diffraction, non pris en compte ici, on choisira $d_3$
de mani{\`e}re {\`a} v{\'e}rifier largement cette in{\'e}galit{\'e}.

Enfin, l'{\'e}paisseur commune $e$, seul param{\`e}tre restant, est
d{\'e}termin{\'e}e de mani{\`e}re {\`a} ce que le diam{\`e}tre $d_1$ de l'ouverture
d'entr{\'e}e d{\'e}limite un champ stellaire de taille convenable {\`a} fixer.

          \section{Les r{\'e}flexions secondaires}
          \label{sec:Refl}

\begin{figure}[!h]
\begin{center}
\epsfig{file=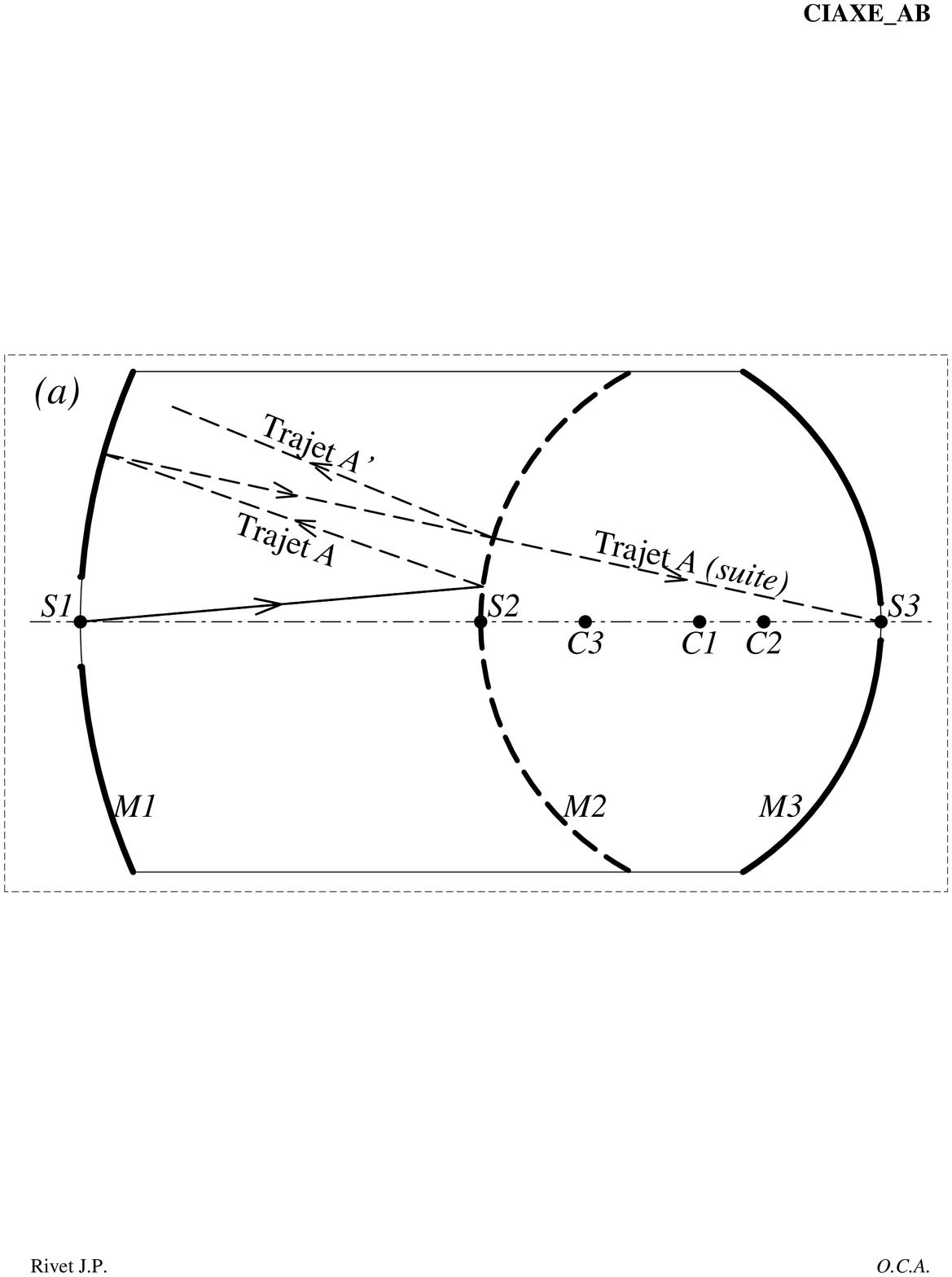,width=6cm,angle=0,clip=}
\epsfig{file=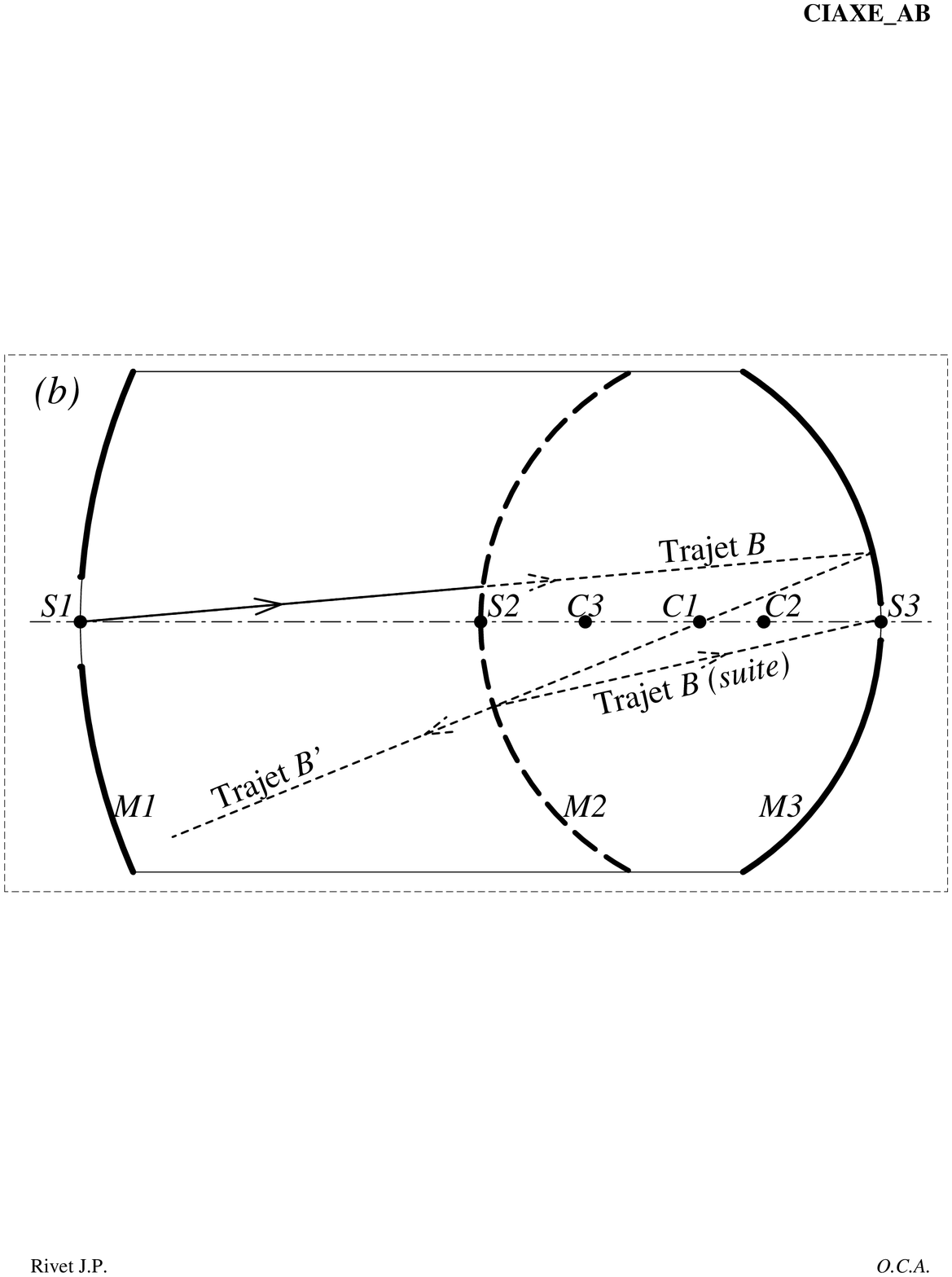,width=6cm,angle=0,clip=}
\end{center}
\caption[Les trajets secondaires $A'$ et $B'$]{\small\it Les
trajets secondaires $A'$ et $B'$. {\sl(a)}\,: Au lieu d'{\^e}tre
transmis par la s{\'e}paratrice $M_2$ pour atteindre la sortie $S_3$
comme le rayon $A$, le rayon $A'$ est r{\'e}fl{\'e}chi par $M_2$.
{\sl(b)}\,: Au lieu d'{\^e}tre r{\'e}fl{\'e}chi par la s{\'e}paratrice $M_2$ pour
atteindre la sortie $S_3$ comme le rayon $B$, le rayon $B'$ est
transmis par $M_2$.}
 \label{fig:CIAXE_ABprim}
\end{figure}

Du seul point de vue de l'optique g{\'e}om{\'e}trique, les deux trajets
$A$ et $B$ ne sont pas les seuls possibles, puisque chaque passage
par la s{\'e}paratrice $M_2$ engendre un sous-faisceau transmis et un
sous-faisceau r{\'e}fl{\'e}chi. Il s'en suit une arborescence infinie de
trajets possibles. Par exemple, lorsque le faisceau qui suit le
trajet $A$ aborde pour la \emph{seconde} fois la s{\'e}paratrice
$M_2$, il se divise en deux sous-faisceaux\,:
 \begin{itemize} \itemsep=0pt
 \item un sous-faisceau transmis qui emprunte donc la suite du trajet $A$,
 \item un sous-faisceau r{\'e}fl{\'e}chi qui emprunte ensuite le trajet alternatif $A'$
 (voir Figure~\ref{fig:CIAXE_ABprim}-a). Ce sous-faisceau correspond {\`a} une
 onde sph{\'e}rique qui diverge en direction du miroir $M_1$, en semblant provenir
 du point $C_1$, centre de courbure de $M_1$.
 \end{itemize}
De m{\^e}me, lorsque le faisceau qui suit la trajectoire $B$
aborde pour la \emph{seconde} fois la s{\'e}paratrice $M_2$, il se
divise lui aussi en deux sous-faisceaux\,:
 \begin{itemize} \itemsep=0pt
 \item un sous-faisceau r{\'e}fl{\'e}chi qui emprunte donc la suite du trajet $B$,
 \item un sous-faisceau transmis qui emprunte ensuite le trajet alternatif $B'$
 (voir Figure~\ref{fig:CIAXE_ABprim}-b). Ce sous-faisceau correspond
 aussi {\`a} une onde sph{\'e}rique qui diverge en semblant provenir
 du point $C_1$.
 \end{itemize}
On va montrer\footnote{Pour un coronographe id{\'e}alement r{\'e}alis{\'e} et
pour un faisceau incident parfait.} que pour un objet non r{\'e}solu
situ{\'e} sur l'axe du coronographe, la totalit{\'e} de l'{\'e}nergie
incidente est renvoy{\'e}e vers l'entr{\'e}e, au moins dans
l'approximation de l'optique g{\'e}om{\'e}trique. Cela garantit, par
conservation de l'{\'e}nergie, que les autres niveaux de
l'arborescence des sous faisceaux, qui existent formellement du
point de vue g{\'e}om{\'e}trique, ne transportent aucune {\'e}nergie vers
l'ouverture de sortie.

Convenons d'appeler $r=\sqrt{R}e^{i\rho}$ et $t=\sqrt{T}e^{i\tau}$
les coefficients de r{\'e}flexion et de transmission de la s{\'e}paratrice
$M_2$ pour l'amplitude du champ. Cette s{\'e}paratrice {\'e}tant suppos{\'e}e
sym{\'e}trique et non absorbante, les modules et arguments des
coefficients de r{\'e}flexion et de transmission doivent v{\'e}rifier
\cite{LLT95a}\,:
 \begin{equation}
   R+T=1\quad\hbox{et}\quad \rho-\tau={{\pi}\over{2}}+k\,\pi,
 \label{eq:RT}
 \end{equation}
o{\`u} $k$ est un entier arbitraire.

 Pour une amplitude incidente
{\'e}gale {\`a} l'unit{\'e}, l'onde sph{\'e}rique divergente issue du trajet $A'$
porte une amplitude $r^2$, et l'onde sph{\'e}rique divergente issue du
trajet $B'$ porte une amplitude $-t^2$. Le signe n{\'e}gatif vient du
d{\'e}phasage de $\pi$ par passage au point focal r{\'e}el $C_1$
\cite{G90a,BW99a}. La somme de ces deux ondes est toujours une
onde sph{\'e}rique divergente, semblant provenir de $C_1$, et qui
porte une amplitude $r^2-t^2$. Compte tenu des
relations~(\ref{eq:RT}) sur les modules et arguments des
coefficients $r$ et $t$, cette somme vaut $e^{2i\rho}$, dont le
module au carr{\'e} vaut $1$. En d'autres termes, cette onde sph{\'e}rique
porte la totalit{\'e} de l'{\'e}nergie incidente. De plus, le point
d'origine (apparent) de cette onde {\'e}tant le centre de courbure
$C_1$ de la surface $M_1$, ce dernier renvoie l'onde en question
sur elle-m{\^e}me. Le principe du retour inverse de la lumi{\`e}re
garantit alors que cette onde qui porte toute l'{\'e}nergie incidente
est renvoy{\'e}e sur ses pas, c'est-{\`a}-dire vers l'entr{\'e}e $S_1$.

En cons{\'e}quence, la lumi{\`e}re incidente venant d'une source non
r{\'e}solue situ{\'e}e sur l'axe du dispositif est int{\'e}gralement renvoy{\'e}e
vers l'entr{\'e}e (du moins dans le cadre de l'optique g{\'e}om{\'e}trique
paraxiale). Par conservation de l'{\'e}nergie, aucune {\'e}nergie
n'atteint la sortie du dispositif.

 Pour {\^e}tre pr{\'e}cis, il faut incorporer au raisonnement ci-dessus
le fait que les surfaces  $M_1$ et $M_3$ ne sont r{\'e}fl{\'e}chissantes
qu'{\`a} l'ext{\'e}rieur des ouvertures d'entr{\'e}e et de sortie. Or, les
diverses surfaces d'onde qui interviennent dans le probl{\`e}me sont
limit{\'e}es ext{\'e}rieurement par l'ouverture du t{\'e}lescope, et de plus
``perfor{\'e}es'' en leur centre par l'ombre port{\'e}e de l'obstruction
centrale du t{\'e}lescope (si l'on n{\'e}glige la diffraction). Si les
ouvertures d'entr{\'e}e et de sortie ont des diam{\`e}tres suffisamment
faibles pour {\^e}tre dans l'ombre port{\'e}e de cette obstruction
centrale, leur pr{\'e}sence ne modifient pas les conclusions du
raisonnement {\'e}nerg{\'e}tique ci-dessus.

          \section{Conclusion}
          \label{sec:Concl}

La r{\'e}alisation pratique de ce dispositif pose des probl{\`e}mes de
recherche et d{\'e}veloppement qui ne semblent pas insolubles. Un des
modes de r{\'e}alisation {\`a} l'{\'e}tude est l'usinage au diamant sur un
tour nanom{\'e}trique, dont les performances affich{\'e}es paraissent
compatibles avec les tol{\'e}rances exig{\'e}es pour la r{\'e}alisation. Le
mat{\'e}riau optique pourrait {\^e}tre le $ZnSe$, mat{\'e}riau transparent
de $0,5\,\mu m$ {\`a} $20\,\mu m$, dont l'indice est voisin de $1+\sqrt{2}$
\textcolor{Modif}{en bande $K$},                                                       %@@@@6
%pour les longueurs d'onde utiles
ce qui permet la r{\'e}alisation d'une s{\'e}paratrice {\'e}quilibr{\'e}e sans traitement.
\textcolor{Modif}{Les effets de chromatisme et de polarisation pour des incidences     %@@@@5,6
non nulles au niveau de la s{\'e}paratrice restent minimes. La d{\'e}gradation li{\'e}e {\`a} ces
seuls effets permet un taux de r{\'e}jection sup{\'e}rieur {\`a} $10^5$ entre $2\,\mu m$ et
$3\,\mu m$, pour un faisceau d'entr{\'e}e ouvert {\`a} $F/15$.}

 Le dispositif pr{\'e}sent{\'e} semble {\^e}tre une voie
prometteuse pour disposer de moyens d'imagerie {\`a} haute dynamique
par coronographie. Il allie les performances coronographiques du
CIA classique (achromaticit{\'e}, pas de discontinuit{\'e} de la surface
d'onde, visibilit{\'e} de $50\,\%$ jusqu'{\`a} $1/3$ du rayon d'Airy; voir
\cite{BRG00a}), avec la facilit{\'e} d'insertion et d'utilisation d'un
syst{\`e}me compact, coaxial et verrouill{\'e} en diff{\'e}rence de marche.

% Les remerciements sont dans une section, sans num{\`E}rotation
%\section*{Remerciements}
% Remerciements - texte ici

\end{document}